\documentclass[aip,jap]{revtex4-1}
\usepackage{epsfig,amsmath}

\begin{document}


\title{Josephson junction microwave modulators for qubit control} 



\author{O. Naaman}
\email[]{ofer.naaman@ngc.com}
\author{J. A. Strong}
\author{D. G. Ferguson}
\author{J. Egan}
\author{N. Bailey}
\author{R. T. Hinkey}

\affiliation{Northrop Grumman Systems Corp., Baltimore, Maryland 21240, USA }


\date{\today}

\begin{abstract}
We demonstrate Josephson junction based double-balanced mixer and phase shifter circuits operating at 6-10 GHz, and integrate these components to implement both a monolithic amplitude/phase vector modulator and an I/Q quadrature mixer. The devices are actuated by flux signals, dissipate no power on chip, exhibit input saturation powers in excess of 1 nW, and provide cryogenic microwave modulation solutions for integrated control of superconducting qubits.
  
\end{abstract}


\maketitle 


\section{Introduction}\label{introduction}
Control of superconducting qubits has, to date, relied almost exclusively on room-temperature generated signals. While state-of-the-art room temperature control techniques have been tremendously successful,\cite{Kelly15,Corcoles15} the engineering challenges associated with the delivery of high bandwidth microwave signals to the qubit chip, including thermal management, signal integrity,\cite{Gustavsson13} packaging, wiring,\cite{Bejanin16} and device layout, are poised to become key bottlenecks in larger quantum information systems.

A nascent strategy for alleviating the room-to-cryo bandwidth bottleneck is to integrate microwave multiplexing, routing, and modulation circuits in the cryogenic environment alongside the qubits.\cite{Hornibrook15} Several groups have recently demonstrated Josephson junction based amplifers \cite{Mutus13,Macklin15} and circulators,\cite{Sliwa15,Kerchoff15} as well as switches for on-chip routing;\cite{Naaman16,Chapman16,Pechal16} however, there is still a need for microwave modulation technologies \cite{Rafique09,Brummer11} that can meet the dissipation and power requirements associated with integrated qubit control. Here, we describe a double-balanced mixer and a phase-shifter that are built in a superconducting integrated circuit with Josephson junction active elements. We use these components to implement a monolithic Josephson junction vector modulator, as well as an I/Q quadrature modulator\textemdash a device that is used ubiquitously to generate shaped microwave pulses for qubit control. The devices operate in the 6-10 GHz band with no on-chip power dissipation, greater than 1~nW saturation power, greater than 25~dB LO/RF isolation, and with a DC-850~MHz IF bandwidth.

\section{Double-Balanced Mixer}\label{mixer}
The prototypical room-temperature double balanced mixer is built with four diodes arranged in a bridge configuration, with the LO and RF ports of the mixer coupled respectively to the common and balanced modes of the bridge.\cite{PozarBook} When operated as a modulator, an IF signal biases the diodes pairwise to un-balance the bridge so that a positive (negative) IF voltage causes a portion of the LO signal to appear in-phase (180$^\circ$ out of phase) across the RF port, while zero IF voltage leaves the bridge completely balanced and the RF port isolated by symmetry. We implement a superconducting version of the double-balanced mixer by relying on the flux-tunable inductance of Josephson junctions in place of the voltage-tunable resistance of the diodes. To this end, our design must address two challenges that are common to the implementation of microwave devices with Josephson junctions rather than semiconducting components. First, the impedance presented by the superconducting circuit embedding the junctions is typically low and inductive, and requires proper matching to the 50 $\Omega$ environment. We address this challenge by embedding the junctions in a band-pass filter network that takes advantage of the junctions' inductive impedance. Second, operation of Josephson junction devices at the several-GHz frequency range typically calls for junctions with critical currents on the order of $\frac{\hbar\omega_0}{2eZ_0}\sim1 \mu$A (\textit{e.g.} in a $Z_0=20~\Omega$ circuit operating at $\omega_0/2\pi=10$ GHz); this limits the saturation power in these devices to picowatt levels. By using junction arrays in our devices instead of single junctions, we can increase their saturation power to the nanowatt scale, which makes them relevant to qubit control applications. 

\begin{figure}
\includegraphics[width=3.0in]{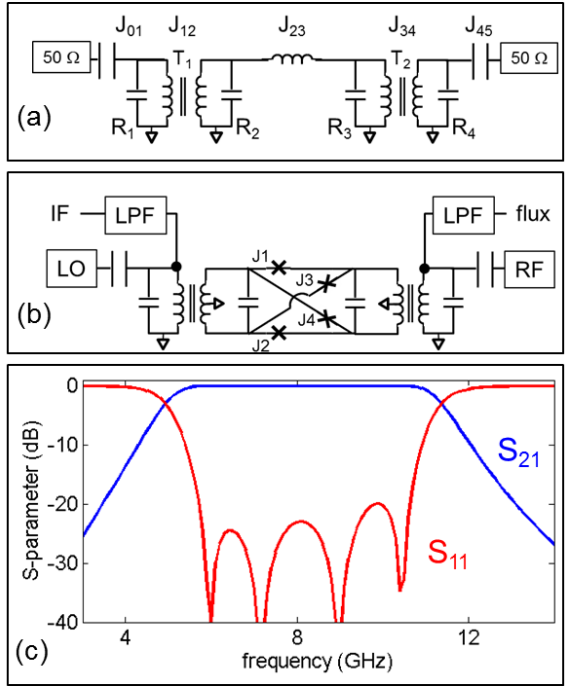}
\caption{\label{fig1}(a) Schematic of a 4-pole filter prototype for embedding a Josephson junction bridge. (b) Schematic of the double-balanced mixer obtained by balancing the circuit in (a) and replacing the inductance of $J_{23}$ with junctions J$_1$ and J$_2$. Junctions J$_3$ and J$_4$ complete the balanced bridge. DC flux bias and IF signal lines are connected as shown via low pass filters (LPF). (c) Simulated 2-port S-parameters of the mixer in its maximum transmission `on' state between the LO and RF ports (port labels are interchangeable by symmetry). The simulation was performed with Agilent ADS, using linear inductances $L_J=460$~ pH for junctions J$_1$ and J$_2$, and $50L_J$ for junctions J$_3$ and J$_4$. Capacitors were simulated as ideal components with values given in the text. Transformers T$_{1,2}$ were first simulated in HFSS from the physical layout, and black-boxed as data-based components in the ADS simulation.}
\end{figure}

To design the Josephson double-balanced mixer we first construct a coupled-resonator band-pass embedding network following the procedure outlined in Ref.~\onlinecite{Naaman16}. We use a 4-pole network with Chebychev response, a center frequency of $\omega_0/2\pi$=8 GHz, and a target bandwidth of 4 GHz. The resulting network, shown in Fig.~\ref{fig1}(a), has four parallel LC resonators R$_1$-R$_4$, coupled with admittance inverters $\{J_{ij}\}$ whose values, having units of $1/\Omega$, are calculated from the filter prototype coefficients $\{g_i\}$.\cite{MYJ} We implement the first and last inverters, $J_{01}$ and $J_{45}$, using capacitors as in Ref.~\onlinecite{Naaman16}, and use inductive transformers T$_1$ and T$_2$ to implement inverters $J_{12}$ and $J_{34}$. The remaining inverter, $J_{23}$, is inductive and can be replaced by a Josephson junction whose critical current is set to $I_c=\frac{\hbar}{2e}\omega_0J_{23}$.

The circuit of Fig.~\ref{fig1}(a), with a Josephson junction inserted for $J_{23}$ is functionally similar to the microwave switch described in Ref.~\onlinecite{Naaman16}. Here, however, the transformer coupling of the filter's inner section via T$_{1,2}$ allows us to balance this section, as shown in Fig.~\ref{fig1}(b), with junction J$_2$ balancing the inductance of junction J$_1$. To complete a balanced bridge configuration, we add the junctions J$_3$ and J$_4$ as shown in the figure. The inner section of the filter, containing resonators R$_2$, R$_3$, and the junction bridge could be left floating, but we chose to ground the center tap of the resonators' inductors to short out parasitic common modes from propagating through the device. Unlike diodes, Josephson junctions do not possess an inherent directionality and the circuit has to be biased with a DC flux, which we apply via transformer T$_2$ by supplying current to the inductor of resonator R$_4$ through a low-pass filter. Finally, the IF signal can be similarly applied via transformer T$_1$ by driving the inductor of resonator R$_1$. Since both DC and IF signals are provided as flux via superconducting transformers, this and all other devices reported here could in principle be controlled and actuated by on-chip superconducting drivers \cite{Semenov15,Naaman14} backed by energy efficient single flux quantum  logic.\cite{Herr11,Mukhanov11,Holmes13} 

The double-balanced Josephson junction mixer operates as follows. When the polarity of the flux induced in the bridge by the IF current is the same as that induced by the DC flux bias, the resulting circulating currents cancel on junctions J$_1$ and J$_2$ while adding up on junctions J$_3$ and J$_4$. This establishes a low-inductance, impedance matched direct signal path through J$_1$ and  J$_2$, while junctions J$_3$ and J$_4$ are in a high inductance state which suppresses the transmission along the crossed path. In this state, which we call the `on' state, the LO signal propagates directly to the RF port. When the polarity of the IF current is opposite to that of the DC flux bias, in what we call the `inverted' state, the induced currents sum on J$_1$ and J$_2$ instead, leaving the crossed path through J$_3$ and J$_4$ matched. In this state the LO signal propagates to the RF port with an additional 180 degree phase shift. When the IF current is zero, in the `off' state, the bridge is balanced and no signal propagates to the output.

Fig.~\ref{fig1}(c) shows an S-parameter simulation of our mixer in its on state, with near unity signal transmission through the device over the design frequency band and better than 20 dB return loss. We used HFSS to extract the S-parameters of transformers T$_{1,2}$ from the layout of the device in order to capture any parasitic capacitance between the transformer coils, and fine-tuned the other circuit elements accordingly to recover the approximate equi-ripple response shown in the figure. The design value of the coupling capacitors (inverters $J_{01}$ and $J_{45}$) is 0.818 pF, the capacitance of the resonators R$_1$ and R$_4$ is 0.343 pF, and that of resonators R$_2$ and R$_3$ is 0.925 pF. The transformers T$_{1,2}$ were designed to have a primary inductance of 1 nH, secondary inductance of 840 pH, and a mutual inductance of 550 pH. The Josephson inductance of the junctions is $L_J$ = 460 pH. 

\begin{figure}
\includegraphics[width=3.37in]{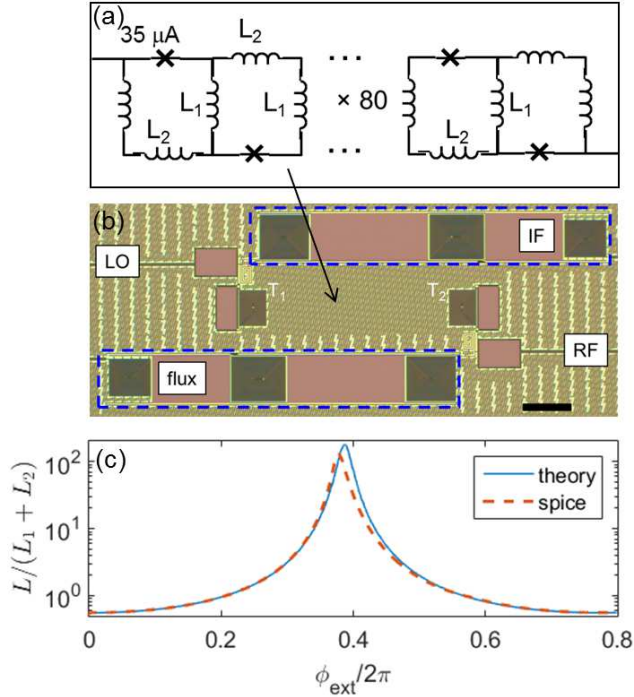}
\caption{\label{fig2}(a) Schematic of the junction array used as the active element in the mixer. Each of the $N=80$ stages contains one $I_c=35~\mu$A junction shunted by a segment of the inductive spine with $L_1=1.457$~ pH and $L_2=3.525$~ pH. (b) Optical micrograph of the mixer corresponding to Fig.~\ref{fig1}(b). The junction array bridge is underneath the ground plane and is not visible. Dashed boxes highlight the low-pass filters on the IF and DC flux ports. Scale bar is 50 $\mu$m. (c) The inductance per stage of the junction array shown in (a) vs applied phase bias, as calculated from Eq.~(\ref{Ltot}) (solid), and simulated using WRSpice (dashed).}
\end{figure}

While the mixer can be constructed with only four junctions, the critical current of each of the junctions in the design above would be $I_c = 0.72~\mu$A, limiting the saturation power to P$_\textrm{sat} < -90$ dBm: a sufficient power level in qubit readout applications, but much lower than the $-60$ dBm level typically desired for qubit control. A common method for increasing the saturation power of Josephson devices is to replace the single junction with a series array of $N$ junctions, each having a critical current of $NI_c$.\cite{Eichler14} This arrangement, however, is susceptible to phase slips\cite{Pop10} and cannot sustain the relatively large phase bias, a significant fraction of $\Phi_0/2$ per junction, required to operate the mixer. To stabilize a junction array against phase slips, we use the configuration shown in Fig.~\ref{fig2}(a)\textemdash a series of junctions connected across a meandering inductive spine, reminiscent of the superinductor design of Ref.~\onlinecite{Bell12}. Our array contains 80 Josephson junctions, each having a critical current of $I_c = 35~\mu$A, and each inductively shunted by a segment of the spine with $L_1 = 1.457$ pH and $L_2 = 3.525$ pH. The inductive spine precludes flux from crossing the array, and the linear inductance shunting each of the junctions is sufficiently small, $2L_1+L_2<L_J$, to make the individual loops\textemdash essentially low-inductance rf-SQUIDs\textemdash mono-stable for all phase bias. The array, therefore, does not have a lower energy state that can be reached by a phase slip event. In total our mixer, with each of J$_1$-J$_4$ replaced by an 80-junction array, contains 320 junctions.

If the array is sufficiently long so that edge effects can be neglected, we can approximate its inductance by assuming translational invariance \cite{Bell12} and finding the equilibrium phase drop across each of the junctions, $\delta_0$, as a function of an applied phase bias, $\phi_{\textrm{ext}}$, across each stage of the array:
\begin{equation}\label{equilib}
  \delta_0\left(\frac{1}{L_1}+\frac{1}{L_2}\right)+\frac{1}{L_J}\sin\delta_0=\phi_{\textrm{ext}}\left(\frac{1}{L_1}+\frac{2}{L_2}\right).
\end{equation}
The total inductance of the array is then found by:
\begin{equation}\label{derivs}  L\left(N\phi_{\textrm{ext}}\right)=N\left(\frac{\hbar}{2e}\right)^2\times\left(\frac{\partial^2u}{\partial\phi_{\textrm{ext}}^2}+\frac{\partial^2u}{\partial\phi_{\textrm{ext}}\partial\delta_0}\frac{d\delta_0}{d\phi_{\textrm{ext}}}\right)^{-1},
\end{equation}
where $u$ is the potential energy per array stage. Evaluating the derivatives in Eq.~(\ref{derivs}), we obtain:
\begin{equation}\label{Ltot}
L\left(N\phi_{\textrm{ext}}\right)=N\frac{\left(L_1+L_2\right)L_J+L_1L_2\cos\delta_0}{L_J+\left(4L_1+L_2\right)\cos\delta_0}.
\end{equation}

Fig.~\ref{fig2}(c) shows the inductance per stage of the array, normalized to the linear inductance per stage $L_1+L_2$, on a semilog plot vs the applied phase bias per stage. The solid line in the figure is calculated using Eq.~(\ref{Ltot}), and is compared to the inductance extracted from a WRSpice transient simulation \cite{Whiteley91} (dashed), in which we applied flux to a loop containing the 80-junction array and calculated the derivative of the loop current with respect to flux. The simulation is in reasonable agreement with theory, and both show a dramatic 100-fold increase of effective array inductance when the phase bias is around $0.38\times2\pi$ radians per stage. Our devices require the array inductance to modulate by a factor of at least $\sim50$, and both theory and simulations indicate that this is readily achievable.

Our devices were fabricated by D-Wave Systems, Inc.~in a 6-layer Nb process \cite{Johnson10} with Nb/Al/AlO$_x$/Nb trilayer junctions having a critical current density of 10 kA/cm$^2$. In this process, the 35 $\mu$A array junctions are circular with a diameter of 0.67 $\mu$m. Fig.~\ref{fig2}(b) shows an optical micrograph of the mixer; the Josephson junction array bridge is located underneath the ground-plane and is not visible. The IF and DC flux lines are connected to the device via on-chip 1 GHz singly-terminated low-pass filters, which are highlighted in the figure with dashed boxes. In addition to the mixer, each chip contained a $50~\Omega$ through line for calibration purposes and a passive mock-up of the mixer with linear inductances in place of the junctions, as well as a phase shifter and an amplitude/phase vector modulator that we discuss below. We have measured a total of four chips from two different wafers with comparable results. All measurements were performed at 4.2 K in liquid helium.

\begin{figure*}[t]
\includegraphics[width=5in]{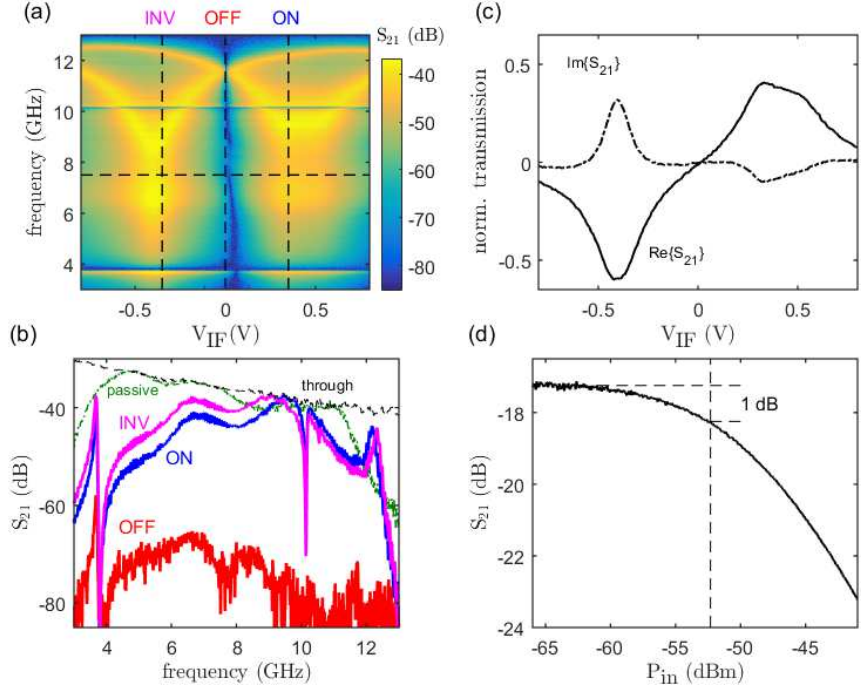}
\caption{\label{fig3}(a) Raw transmission, $S_{21}$, from the LO to the RF port of the mixer, vs frequency and IF bias. A current of 0.7 mA was applied to the DC flux port of the mixer throughout the experiment. The mixer's on (ON), off (OFF), and inverted (INV) states are indicated as vertical dashed lines at $V_{\textrm{ON}}=0.35$ V, $V_{\textrm{OFF}}=0$ V, and $V_{\textrm{INV}} = -0.35$ V, applied across a 1 k$\Omega$ resistor at room temperature. $S_{21}$ vs frequency at these bias points are shown in panel (b), along with the transmission of an on-chip 50 $\Omega$ through calibration line (`through', dashed), and a passive mock-up of the device (`passive', dotted). (c) Real and imaginary parts of $S_{21}$ vs IF bias, normalized to the through line transmission, taken along the horizontal dashed line in panel (a) at 7.5 GHz. (d) Raw  $S_{21}$ at 6.85 GHz and V$_{\textrm{IF}}=-0.41$ V vs input LO power. For this measurement a 20 dB attenuator was removed from the input signal chain.}
\end{figure*}
We first characterize the operation of the mixer under static bias conditions. Fig.~\ref{fig3}(a) shows the transmission, $S_{21}$, of the mixer as a function of frequency and IF voltage as applied to the chip via a room-temperature 1 k$\Omega$ resistor. The input power to the device was $-76$ dBm; we observed no difference in the device behavior at lower input powers. The raw data in the figure represents $S_{21}$ at the reference plane of the network analyzer, and includes a $-50$ dB fixed attenuator at the device's input, a +26 dB amplifier at its output, and approximately 12 dB of round-trip cable losses. A DC current of 0.7 mA was applied to the mixer's flux bias port throughout the experiment. The figure shows two regions of high transmission in the 6-10 GHz range, symmetrically positioned in IF voltage around a deep transmission null at V$_{\textrm{IF}}$=0. We identify the transmission lobe at positive IF voltage as the mixer's on state, and the lobe at negative IF voltage as the mixer's inverted state. In the off state, at V$_{\textrm{IF}}$=0 the carrier's energy is reflected to the input port.

Fig.~\ref{fig3}(b) compares the transmission in the on, off, and inverted states of the mixer, as indicated by dashed lines in panel (a), to an on-chip 50 $\Omega$ through line. The $S_{21}$ traces in the on and inverted states follow band-pass filter characteristics with at most 10 dB insertion loss in the 6-10 GHz range, and the on/off ratio is greater than 25 dB over the same frequency range. The asymmetry of the response in the on and inverted states is reproducible in all devices measured and reflects an inherent asymmetry in the layout of the mixer. The transmission spectra exhibit two strong resonances near 4 and 10 GHz, which can also be seen in panel (a). The frequency of these resonances appear to be independent of IF voltage, and are therefore not likely to be associated with the junctions. The resonances are absent in the transmission through a passive mock-up of the device, which did not have the IF and DC flux bias lines (Fig.~\ref{fig3}(b), green, dotted). This leads us to implicate these resonances, and their impact on the mixer insertion loss, on parasitic modes of the large low-pass filters on these lines. Panel (c) in the figure shows a slice through the data of panel (a) at a constant frequency of 7.5 GHz, where both real and imaginary parts (to within an arbitrary global phase) of the transmission, normalized to the through calibration data, are plotted vs V$_{\textrm{IF}}$; the data demonstrate the inversion of the transmitted signal at negative V$_{\textrm{IF}}$. Both quadratures appear to modulate with V$_{\textrm{IF}}$ due to the reactive nature of the active elements of the mixer, a behavior we also observed in simulations. On a polar plot, $S_{21}$(V$_{\textrm{IF}}$) would trace an S-like curve passing through the origin.

To characterize the power dependence of the device, we performed CW-frequency power-swept $S_{21}$ measurements as a function of IF bias.  We observe that the mixer's on and inverted states, where the nonlinearity of two of the junction arrays is strongest, are first to saturate with increasing power. A slice of the data at V$_{\textrm{IF}}= -0.41$ V (inverted state) and $\omega/2\pi$=6.85 GHz is shown in panel (d), indicating that the 1 dB saturation power of this device is approximately P$_{\textrm{1dB}} = -53$ dBm at the reference plane of the chip.
\begin{figure}
\includegraphics[width=3.37in]{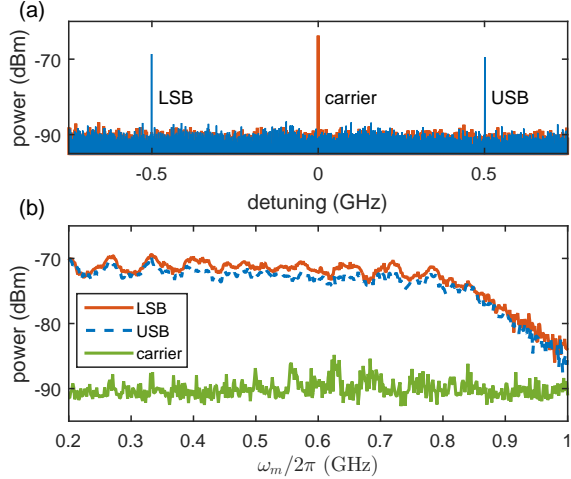}
\caption{\label{fig4}(a) Spectrum analyzer traces of the Josephson double balanced mixer RF output vs detuning $\Delta=(\omega-\omega_{\textrm{LO}})/2\pi$. With 0.36 mA dc current bias to the IF port, the mixer is in its on state and the $\omega_{\textrm{LO}}/2\pi=7.5$~GHz, un-modulated carrier propagates to the output (red). When an ac modulating signal is applied to the IF port, $\omega_m/2\pi=500$~MHz (blue), the carrier is suppressed and only sidebands are visible in the spectrum at $\Delta=\pm\omega_m/2\pi$. (b) Magnitude of the carrier ($\omega_{\textrm{LO}}/2\pi=7.5$~GHz), lower sideband (LSB, $\frac{\omega_{\textrm{LO}}-\omega_m}{2\pi}$), and upper sideband (USB, $\frac{\omega_{\textrm{LO}}+\omega_m}{2\pi}$) in a carrier-suppressed modulation experiment, as a function of frequency $\omega_m$ of the modulating signal driving the IF port of the mixer.}
\end{figure}

Next, we characterize the response of the mixer to a sinusoidal signal applied to its IF port. We fed a 7.5 GHz, $-76$ dBm carrier tone to the mixer LO port and monitored the output from its RF port with a spectrum analyzer; the spectra are shown in Fig.~\ref{fig4}(a) as a function detuning from the carrier frequency. The first spectrum (red in the figure) was taken with the mixer held in its on state and with no modulation applied. The second spectrum (blue) was taken with a $-30$ dBm, 500 MHz tone driving the IF port. The spectrum shows two sidebands at the modulation frequency, while the carrier is suppressed to below the noise floor of the measurement. This carrier-suppressed modulation confirms that the modulation is balanced: the carrier modulates through zero and inverts in the negative portions of the IF cycle, as also illustrated in Fig.~\ref{fig3}(c). When the carrier power is increased beyond approximately $-76$~dBm we start observing spurious sidebands at multiples of the IF frequency; we have not, however, characterized the mixer's nonlinearity or intermodulation products. In Fig.~\ref{fig4}(b) we trace the magnitude of the upper (blue) and lower (red) sidebands, as well as that of the carrier (green), as we vary the frequency $\omega_m/2\pi$ of the IF signal from 200 MHz to 1 GHz. The data shows that the carrier remains suppressed throughout the whole modulation frequency range, and that the magnitude of the sidebands rolls off with a $-3$ dB point at 850 MHz, somewhat lower than the designed 1 GHz cut-off frequency of the on-chip low-pass filter on the IF port.

Because the mixer is non-dissipative and has no gain, it is not expected to contribute its own noise to the signal. However, noise associated with the DC flux and IF control lines will, in general, result in both amplitude- and phase-modulation noise imprinted on the RF output. From the data in Fig.~\ref{fig3}(c) we can calculate the mixer sensitivity to control noise, and if the DC and IF lines are both matched and thermalized to 4 K then we expect both amplitude and phase noise power density to be less than approximately $-157$~dBc/Hz, referenced to the LO input power. Flux noise in the balanced bridge and critical current noise in the junctions will have an additional contribution to the overall modulation noise on the signal. 

\section{Phase Shifter}\label{phase}
\begin{figure*}
\includegraphics[width=5in]{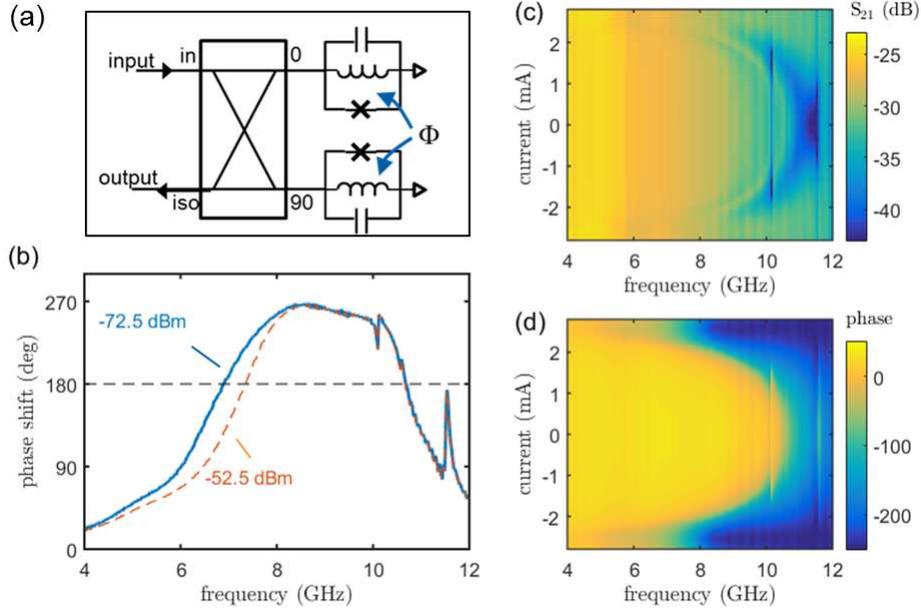}
\caption{\label{fig5}(a) Schematic of the phase shifter, composed of a 90-degree hybrid and two tunable resonators, each containing a 66-junction array marked by the junction symbol. A phase control current coupled inductively to the resonators applies flux to the two tunable resonators in tandem. (b) Maximum phase shift vs frequency at input powers of $-72.5$~dBm (blue, solid) and $-52.5$~dBm (red, dashed), calculated as the phase of the ouput signal at phase control current of 2.48 mA relative to that at 0 mA. (c) Magnitude and (d) phase of the transmission $S_{21}$ between the input and output ports in panel (a) vs frequency and as a function of phase control current, for an input power of $-72.5$~dBm.}
\end{figure*}

Having demonstrated the operation of a Josephson junction double-balanced mixer, we continue by describing a Josephson junction based phase shifter, a second component that we used in our implementation of a vector modulator. A schematic of the phase shifter is presented in Fig.~\ref{fig5}(a). Two over-coupled flux-tunable LC resonators, each containing a 66-junction array similar to Fig.~\ref{fig2}(a) (indicated by a junction symbol in the schematic), are connected to the $0^\circ$ and $90^\circ$ ports of an on-chip coupled-line 90-degree hybrid. The input signal splits evenly between the two arms of the hybrid, reflects off of the two resonators, and re-combines constructively at the `isolated' port of the hybrid, which serves as the output port for the device. If the reflections off of the two resonators have the same magnitude and phase, none of the reflected power reaches the input port, resulting in unity transmission between the input and the output ports. By applying flux to the two resonators in tandem we change their frequency with respect to that of the input signal and therefore the phase of the reflected power. We designed the resonators with capacitances and inductances of 1.48 pF and 820 pH respectively, and the inductance of the junction array at zero flux was set to 205 pH.

Fig.~\ref{fig5}(b) shows the maximum phase shift as a function of frequency, measured from the device's S-parameters at the extrema of the resonators' tunability range. At input powers up to $-72.5$ dBm, the phase can be modulated by more than 180 degrees over a frequency band from 6.9 GHz to 10.7 GHz, with a maximum phase shift of 267 degrees at 8.56 GHz. At higher input powers we observe a degradation of the tunability range at the lowest frequencies, where the junction array nonlinearity is greatest. Panels (c) and (d) show the magnitude and phase, respectively, of the measured $S_{21}$ as a function of frequency and current in the flux bias line to the resonators. The data show that while there is a large change in the phase of the signal as a function of flux bias, there is very little change in the magnitude of the transmitted signal in the 6-10 GHz range, indicating that the resonators are well matched in their frequencies and response to flux, and that there is minimal magnitude and phase imbalance in the on-chip 90-degree hybrid.

\begin{figure}[b]
\includegraphics[width=3.2in]{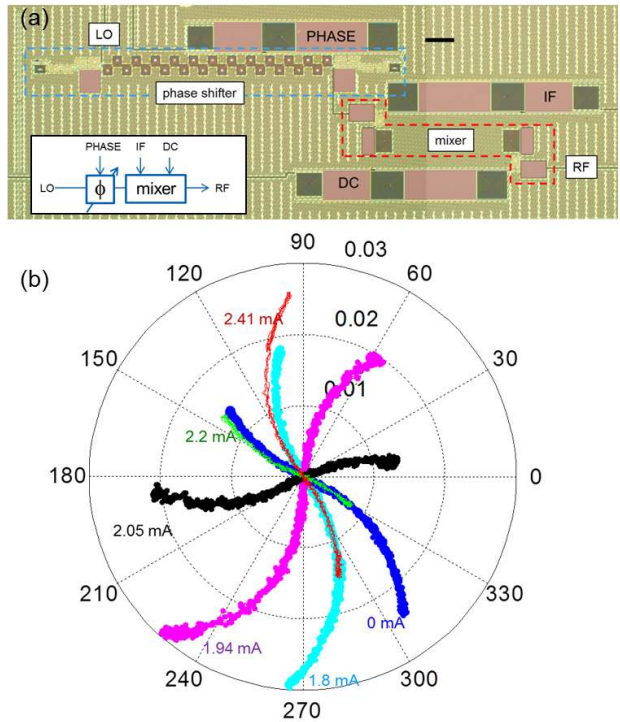}
\caption{\label{fig6}(a) Optical micrograph and block diagram (inset) of the monolithic amplitude/phase vector modulator, scale bar is 50 $\mu$m. The low-pass filters on the PHASE, IF, and DC control lines are labeled correspondingly. (b) polar plot of the complex transmission, $S_{21}$, from the LO to the RF port vs V$_{\textrm{IF}}$ at 7.759 GHz and varying phase control current from $0-2.41$~mA.}
\end{figure}

\section{Microwave Modulators}\label{modulators}
The device shown in Fig.~\ref{fig6}(a) is a monolithic vector modulator, constructed by concatenating the phase shifter described above and the balanced mixer of Fig.~\ref{fig2}(b). The device is controlled by an IF flux that modulates the amplitude of the carrier, and a phase-control flux that modulates the carrier phase, as shown in the inset. The amplitude/phase vector modulator can ideally have zero conversion loss, as compared to a standard I/Q quadrature modulator that must have at least 6 dB of conversion loss, a potential advantage in application where signal losses must be minimized. In Fig.~\ref{fig6}(b) we demonstrate that with this device, the carrier can be modulated with a full coverage of the complex $S_{21}$ plane. The figure shows a series of traces on a polar plot, each trace corresponds to a different phase control current, from 0 to 2.41 mA. At each phase control current we measured the complex $S_{21}$ through the device at 7.759 GHz while sweeping V$_{\textrm{IF}}$ of the mixer from its inverted state, through zero, and to its on state. We see that as the phase control current is increased, the $S_{21}$(V$_{\textrm{IF}}$) traces rotate in the plane until at 2.2 mA  they have rotated by 180 degrees and every point on the complex plane has been visited.  The variation of the maximum magnitude of $S_{21}$ as a function of phase control current is at most 3 dB, and is likely due to mismatches between the output of the phase shifter and the input of the mixer.

\begin{figure}
\includegraphics[width=3.37in]{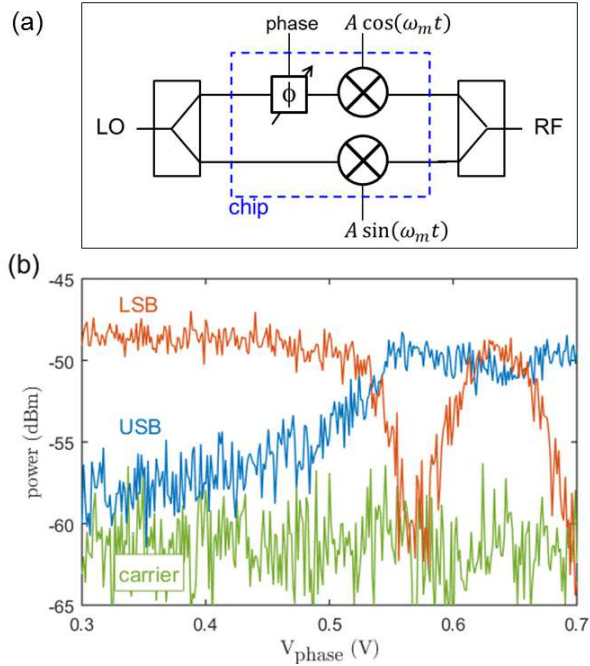}
\caption{\label{fig7} Single sideband, carrier-suppressed modulation of a 7.5 GHz carrier with 200 MHz IF signals. (a) Schematic of the I/Q quadrature modulator and experimental setup. (b) The power at the RF port at the lower sideband (LSB, 7.3 GHz), upper sideband (USB, 7.7 GHz), and carrier (7.5 GHz) frequencies, measured with a spectrum analyzer, and plotted as a function of phase control voltage.}
\end{figure}

In a final experiment, we used the vector modulator, Fig.~\ref{fig6}(a), and the balanced mixer, Fig.~\ref{fig2}(b), co-located on the same chip, together with a pair of room-temperature Wilkinson power splitters/combiners to implement an I/Q quadrature modulator as shown in Fig.~\ref{fig7}(a). The phase shifter portion of the vector modulator, controlled here by a DC voltage source in series with a 30 dB attenuator, was used in this experiment to adjust the relative phase between the 7.5 GHz LO signals feeding the mixers. We drove the I and Q ports of the chip with $\omega_m/2\pi = 200$ MHz, $-30$ dBm sinusoidal signals using two signal generators with a fixed relative phase of 90 degrees, and monitored the output of the modulator using a spectrum analyzer. The 90 degree phase relation between the I and Q baseband signals allows us to perform single-sideband modulation of the carrier: we can select the lower sideband or the upper sideband by setting the relative LO phase to $+90$ or $-90$ degrees, respectively, using the on-chip phase control.

Fig.~\ref{fig7}(b) shows the power at the carrier, upper- and lower-sideband frequencies as a function of phase control voltage, V$_{\textrm{phase}}$. The data show that the carrier is suppressed throughout the phase tuning range, and that the power in the two sidebands changes as a function of V$_{\textrm{phase}}$. For V$_{\textrm{phase}}\leq 0.3$ V, the upper sideband is suppressed and only the lower sideband is visible in the spectrum above the noise floor of the measurement. This indicates an overall 90 degree phase relation between the split LO components as they enter the two mixers. As V$_{\textrm{phase}}$ is increased, the upper sideband grows in magnitude while the lower sideband's magnitude decreases, until at V$_{\textrm{phase}} = 0.53$ V their magnitude becomes equal and 3 dB less than the maximum sideband power: at this point the split LO signals enter the two mixers in phase. At V$_{\textrm{phase}} = 0.57$ V, the LO signals have a relative phase of $-90$ degrees, resulting in the rejection of the lower sideband and only the upper sideband is visible in the spectrum. Increasing V$_{\textrm{phase}}$ further, up to a maximum phase shift at V$_{\textrm{phase}} = 0.63$ V causes the lower sideband to re-appear. This single-sideband carrier-suppressed modulation experiment demonstrates the functionality of the I/Q quadrature modulator and the potential for a drop-in replacement of the standard room-temperature I/Q mixer by a superconducting, Josephson junction based device. The availability of an integrated phase-trim control in the modulator is important in qubit control applications where the orthogonality of qubit rotation operations about the X and Y axes, as determined by the I and Q quadratures of the modulated signal, must be precise.

To summarize, we have designed and demonstrated a double-balanced mixer and a phase shifter built in a superconducting integrated circuit with Josephson junction active elements. We further demonstrated a monolithic vector modulator and an I/Q quadrature mixer built with these components. These devices operate with no on-chip power dissipation, 4 GHz LO and RF bandwidth centered at 8 GHz, and DC-850 MHz IF bandwidth. We have shown that by using Josephson junction arrays in place of single junctions we can significantly increase the saturation power of this type of devices, from the typical picowatt levels common in single junction devices, to the nanowatt level in the devices presented here. These devices provide microwave modulation solutions that operate in the cryogenic environment, can be integrated with or nearby a quantum processor, and join a growing family of microwave switches, amplifiers, and circulators that enables integration of qubit control and readout functionality in the cryogenic space. 




%
%

%



\bibliography{VM_JAP}

\end{document}